\documentclass{iau} 
\usepackage{graphicx}

\title[Early Type Galaxy Halos]{Kinematics and Angular Momentum in Early Type Galaxy Halos} 
\author[Jean P. Brodie, Aaron Romanowsky  and the SLUGGS team] 
{Jean P. Brodie$^1$,
Aaron Romanowsky$^{1, 2}$, and the SLUGGS team$^3$}

\affiliation{$^1$UC Observatories, University of California\\ 1156 High St,
Santa Cruz, CA 95064, USA \\ email: {\tt jbrodie@ucsc.edu} \\[\affilskip]
$^2$ San Jos\'e State University, San Jose, CA \\ email: {\tt aaron.romanowsky@sjsu.edu}\\[\affilskip]
$^3$http://sluggs.ucolick.org}

\pubyear{2015}
\volume{317}  
\jname{The General Assembly of Galaxy Halos: Structure, Origin and Evolution}
\editors{A. Bragaglia, M. Arnaboldi, M. Rejkuba, O. Gerhard} 

\begin{document}

\maketitle

\begin{abstract}
We use the kinematics of discrete tracers, primarily globular clusters (GCs) and planetary nebulae (PNe), along with measurements of the integrated starlight to explore the assembly histories of early type galaxies. Data for GCs and stars are taken from the SLUGGS wide field, 2-dimensional, chemo-dynamical survey (\cite[Brodie et al.\ 2014]{Brodie14}). Data for PNe are from the PN.S survey (see contributions by Gerhard and by Arnaboldi, this volume). We find widespread evidence for 2-phase galaxy assembly and intriguing constraints on hierarchical merging under a lambda CDM cosmology. 
\keywords{galaxies: elliptical and lenticular, galaxies: star clusters, galaxies: formation, galaxies: abundances, galaxies: kinematics and dynamics}
\end{abstract}

\firstsection 
\section{Introduction}
It is now generally agreed that galaxies form in two phases.  A widely accepted scenario involves an early phase, occurring at a redshift of 2 or earlier, that produces a
relatively compact nugget that grows over time by continually accreting lower mass satellites in dry minor mergers (e.g., \cite[Oser et al.\ 2010]{Oser10}, \cite[Naab et al.\ 2014]{Naab14}).  Alternatively, large star forming disks may  evolve passively until quenched by process that relate to their densities or velocity dispersions, perhaps also increasing somewhat in size via dry minor mergers (\cite[van Dokkum et al.\ 2015]{vandokkum15}). 
Given that more than 90\% of the total mass and angular momentum of a galaxy lie beyond one effective radius (R$_e$), it stands to reason that testing models for the assembly of galaxy halos will require observations out to large galactocentric radius.  

The SAGES Legacy Unifying Globular Clusters and GalaxieS (SLUGGS) survey  (\cite[Brodie et al.\ 2014]{Brodie14})  uses SUBARU/SuprimeCam imaging and Keck/DEIMOS spectroscopy to generate 2-dimensional metallicity and kinematic data out to $\sim$3 R$_e$ for galaxy starlight and out to $\sim$10 R$_e$ for globular clusters (GCs) in 25 nearby early type galaxies. 
Here we report initial results from the SLUGGS survey, whose observational component is nearing completion. We also include some results from the Planetary Nebula Spectrograph Galaxy Survey (PN.S), which uses planetary nebulae to explore kinematics and dynamics in 33 nearby galaxies (Arnaboldi et al., 2016, in preparation).

Underpinning the use of GCs to unravel the formation histories of galaxies is the fact that GC formation accompanies all the major star forming events in a galaxy's history.  Typically containing 10$^5$ to 10$^6$ stars, GCs are bright enough to allow integrated spectroscopy out to distances in excess of 50 Mpc.  The vast majority of GCs are as old as we can measure them ($>$10 Gyr) and are bright beacons that were ``along for the ride" during all the mergers and acquisitions that have built the galaxies we see today.

GC systems typically divide into two subpopulations, a blue, metal-poor population that appears to trace the build up of galaxy halos, and a red, metal-rich, population that is linked to bulge development. The subpopulations are distinct not only in metallicity (e.g., \cite[Brodie et al.\ 2012]{Brodie12}), but are also kinematically distinct (\cite[Pota et al.\ 2013]{Pota13}). Like red GCs, planetary nebulae appear to be closely linked to the starlight in ETGs (\cite[Coccato et al. 2009]{Coccato09}).

\begin{figure}
\begin{center}
 \includegraphics[width=5in]{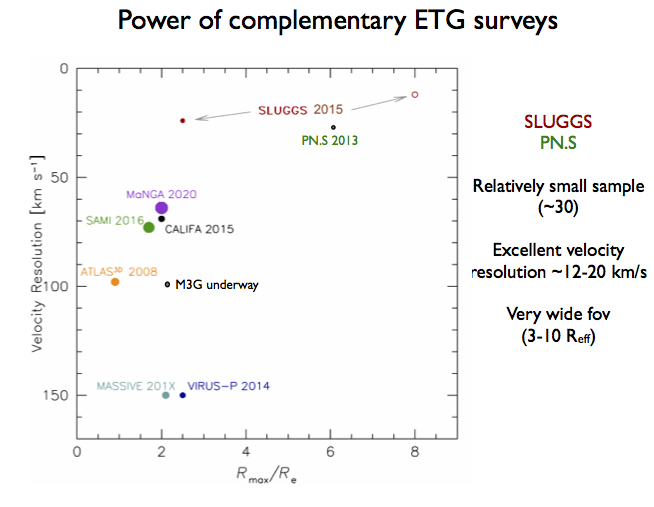} 
 \caption{Complementary ETG surveys. Many other surveys are ongoing that target early type galaxies. While SLUGGS and PN.S provide excellent velocity resolution and wide field coverage, other surveys have much larger galaxy samples.}
   \label{fig1}
\end{center}
\end{figure}

Many surveys are underway that are targeting ETGs. Figure~\ref{fig1} shows a figure of merit for these surveys that is an update of the figure in \cite[Brodie et al. (2014)]{Brodie14}.  SLUGGS and PN.S were designed to offer wide field coverage with very high velocity resolution, but the galaxy sample is relatively small ($\sim$30). Other surveys complement this work by providing the large galaxy samples, albeit with poorer resolution and field coverage. Taken together, these surveys are offering unprecedented insight into the formation histories of ETGs.

\section{Angular Momentum}

\begin{figure}
\begin{center}
 \includegraphics[width=5in]{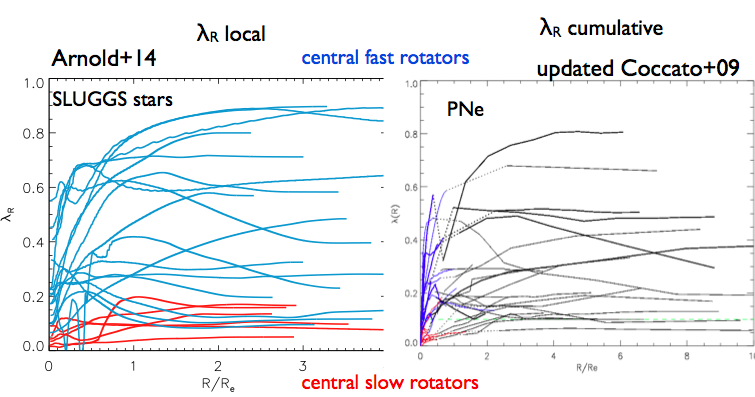} 
 \caption{Specific Angular Momentum. Left panel: local specific angular momentum versus radial extent in units of effective radius for ETG stars from the SLUGGS survey. Right panel: cumulative specific angular momentum versus radial extent for ETGs from the P.NS survey (courtesy of L.Coccato and M.Arnaboldi). Both surveys reveal the same trends. Central slow rotators (red) remain slow. Central fast rotators (blue) may rise, plateau or fall with increasing radius.}
   \label{fig2}
\end{center}
\end{figure}

Figure~\ref{fig2} shows specific angular momentum, $\lambda_R$ versus radial extent for SLUGGS stars and PNe.  The definitions of $\lambda_R$ is different in the two panels. The SLUGGS version uses a local definition defined in successive annuli. The PN.S version is cumulative (as in Atlas$^{3D}$ analyses (\cite[Emsellem et al.\ 2011]{Emsellem11}). We definite a local version to preserve evidence of radial transitions.  Evident from both tracers is the trend for galaxies that were defined as slow rotators based on observations in their central regions, to remain slow with increasing radius.  Centrally defined fast rotators may continue to rise, plateau or decline with increasing radius.  A similar result was obtained form the VIRUS-P survey of 33 massive galaxies (\cite[Raskutti et al.\ 2014]{Raskutti14}).

\section{Velocity Dispersion}

\begin{figure}
\begin{center}
 \includegraphics[width=4in]{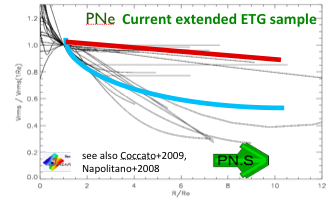} 
 \caption{The radial distribution of V$_{rms}$ for galaxies in the PN.S survey (courtesy of L.Coccato and M.Arnaboldi).}
   \label{fig3}
\end{center}
\end{figure}

\begin{figure}
\begin{center}
 \includegraphics[width=3.4in]{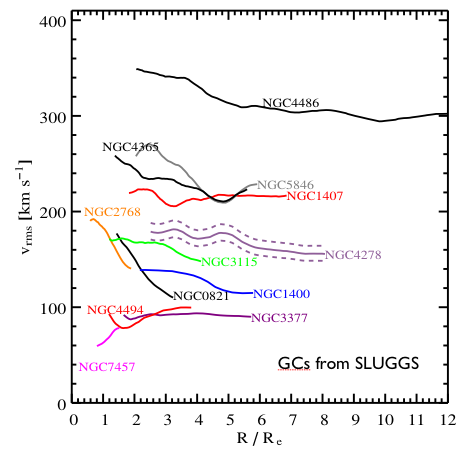} 
 \caption{The radial distribution of V$_{rms}$ of GCs for a subsample of the SLUGGS galaxies.}
   \label{fig4}
\end{center}
\end{figure}

Figure~\ref{fig3} is a plot of root mean square velocity (V$_{rms}$, a proxy for velocity dispersion) against radial extent from PN.S observations of planetary nebulae. The PN.S team find a dichotomy between galaxies displaying nearly flat and steeply declining profiles. Possible explanations for such an effect include a dark matter dichotomy or anisotropy projection effects. See the contribution by Napolitano in this volume for further discussion of this point. In Figure~\ref{fig4} we show the rms velocity as a function of radius for GCs from the SLUGGS survey. Although we do not see a dichotomy in our data, not all of the SLUGGS galaxies have yet been included.

\section{Mass and Dark Matter}

\cite[Pota et al.\ (2015)]{Pota15} carried out multi population dynamical modeling of NGC 1407 using the spherical Jeans equation and employing stars, metal-rich and metal-poor GCs as three independent tracers of the dark matter distribution.  Using a Bayesian MCMC analysis, we determined that different anisotropies are needed to fit the profiles and that the metal-poor GCs have tangential anisotropy. This result for blue (metal-poor) GCs is inconsistent with expectations from hierarchical merging.  Kurtosis measurements for a larger number of SLUGGS galaxies 
(\cite[Pota et al.\ 2013]{pota13}) reveal that the majority of blue GCs are on tangential orbits, while there is a mix of radial and tangential orbits for red (metal-rich) GCs and for PNe.	

\begin{figure}
\begin{center}
 \includegraphics[width=3.4in]{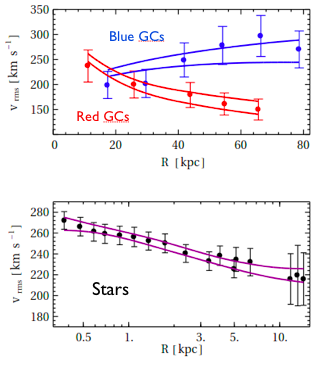} 
 \caption{The best fit to generalized NFW profiles for metal rich and metal poor GCs (top panel) and
 stars (bottom panel). Based on figures in \cite[Pota et al.\ (2015)]{Pota15}. The solid lines are the 1$\sigma$ boundaries of the fits.}
   \label{fig5}
\end{center}
\end{figure}

\begin{figure}
\begin{center}
 \includegraphics[width=3.4in]{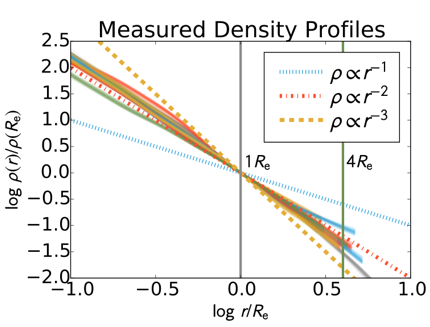} 
 \caption{Measured total density profiles from \cite[Cappellari et al. (2015)]{Capp15}. Solid lines are the individual galaxy measurements which display very little scatter about the isothermal relation with exponent 2.19$\pm$0.04.}
   \label{fig6}
\end{center}
\end{figure}

\cite[Cappellari et al. (2015)]{Capp15} used Jeans axisymmetric models (JAM) on a combination of ATLAS$^{3D}$ and SLUGGS data for 14 galaxies classified as fast rotators based on their central ($<$1R$_e$ kinematics). The JAM models allow spatially varying anisotropy and quite general profiles for the dark matter; no restriction on slope is imposed. This simple axisymmetric model fits the data for all 14 galaxies remarkably well and yields a power law density profile with exponent 2.19$\pm$0.04 over the full range covered by the data  (0.1R$_e$$ >$ r $>$ 4R$_e$; see Figure~\ref{fig6}). The scatter among the 14 galaxies is only 0.14. Since the a power law density profile is not a generic prediction of lambda CDM cosmology, this results offers tight constraints on the cosmological models. It has long been known, e.g., from observations of the gas, that the rotation curves of spiral galaxies flatten at large radius, reflecting the interplay between dark matter and baryons. This is the first indication that the same flattening occurs in early type galaxies revealing a surprising ``dark matter conspiracy" across markedly different galaxy types.

\section{Velocity Position Phase Space}

\begin{figure}
\begin{center}
 \includegraphics[width=3.4in]{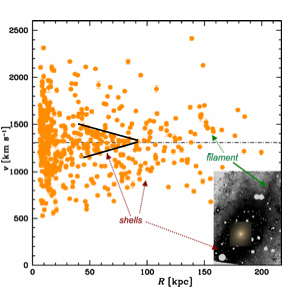} 
 \caption{Velocity-position phase space for GCs around M87 reveals chevron structures that are characteristic of recent massive accretion events. The figure is adapted from \cite[Romanowsky et al. (2011)]{Rom11}. The image in the bottom right hand corner of the plot is from \cite[Mihos et al. (2005)]{mihos05}.}
   \label{fig7}
\end{center}
\end{figure}

Simulations of satellite infall (e.g.\ \cite[Bullock \& Johnston 2005]{Bull05}) show that satellite galaxy accretion can set up a temporary set of nested chevrons in the velocity-position phase space of the accreted material, due to repeated passage near to the center of the more massive (accreting) galaxy.  Direct evidence of this effect was reported by \cite[Romanowsky et al. (2011)]{Rom11},  based on high precision radial velocities of more than 500 GCs associated with M87, the massive elliptical galaxy at the center of the Virgo cluster (Figure~\ref{fig7}).  We inferred that M87 had acquired an L$^*$ galaxy, bringing in $\sim$1000 GCs within the last Gyr. Such characteristic chevrons are erased on a timescale of less than a Gyr, so their presence allows the timing of the accretion event to be estimated.  Recently, the tally of GCs with high precision velocities has risen to more than 1700 and the chevron structure still persists. A chevron is also seen in independently in the phase space distribution of $\sim$300 PNe (\cite[Longobardi et al.\ 2015]{longo15}).

\section{Summary}

Wide field surveys of early type galaxies (such as SLUGGS and PN.S) are revealing widespread evidence in favor of the two-phase paradigm of galaxy assembly. In particular, we find a wide range of stellar and PNe radial profiles. Inner slow rotators remain slow; inner fast rotators can rise, plateau, or fall with increasing radius from the center of the galaxy. Multi-population dynamical modeling using stars, blue and red GCs as 3 independent probes of the gravitational potential provides estimates of DM density, the total enclosed mass, anisotropy, and M/L. We find many instances of unexplained tangential anisotropy for red GCs and PNe, while the majority of the blue GCs have tangential anisotropy. This result is contrary to expectations based on hierarchical merging models. Jeans axisymmetric modeling for stars from a combination of ATLAS$^{3D}$ data for galaxy centers and SLUGGS data for the outer regions (to typically 4R$_e$ ) reveals a remarkable dark matter/baryon conspiracy to produce power law density profiles, with exponent $2.19\pm 0.04$, with very little galaxy-to-galaxy scatter ($<$0.14). Substructure in velocity-position phase space in the distribution of GCs and PNe provides evidence for recent accretion in M87, including an estimate of the mass of the accreted galaxy and event timing.

\section{Acknowledgements}
This work was supported by NSF grant AST-1211995.

\end{document}